\documentclass[aps,tightenlines,nofootinbib,prd]{revtex4}
\pdfoutput=1
\usepackage{amsmath,amscd,amssymb}
\usepackage{color,epsfig}
\usepackage{xfrac}
\usepackage{latexsym}
\usepackage{cancel}
\usepackage{mathrsfs,bigints}
\usepackage{graphicx}

\newcommand{\imu}{{\rm i}}
\newcommand{\zr}[1]{\mbox{\hspace*{#1em}}}
\newcommand{\ID}{\mbox{{\sf 1}\zr{-0.16}\rule{0.04em}{1.55ex}\zr{0.1}}}

\begin{document}

\title{Vacuum Polarization Energy of a Proca Soliton}

\author{Damian A. Petersen$^{a)}$, Herbert Weigel$^{a)}$}

\affiliation{%
$^{a)}$Institute of Theoretical Physics, Physics Department, Stellenbosch University,
Matieland 7602, South Africa}

\begin{abstract}
We study an extended Proca model with one scalar field and one massive 
vector field in one space and one time dimensions. We construct the soliton 
solution and subsequently compute the vacuum polarization energy (VPE) which is 
the leading quantum correction to the classical energy of the soliton. For this
calculation we adopt the spectral methods approach which heavily relies on the 
analytic properties of the Jost function. This function is extracted from the 
interaction of the quantum fluctuations with a background potential generated by 
the soliton.  Particularly we explore eventual non-analytical components that 
may be induced by mass gaps and the unconventional normalization for the 
longitudinal component of the vector field fluctuations. By numerical simulation 
we verify that these obstacles do actually not arise and that the real and imaginary 
momentum formulations of the VPE yield equal results. The Born approximation to the 
Jost function is crucial when implementing standard renormalization conditions. 
In this context we solve problems arising from the Born approximation being 
imaginary for real momenta associated with energies in the mass gap.

\end{abstract}

\maketitle


\section{Introduction}
\label{sec:intro}
We consider solitons as static finite energy solutions to non-linear field
equations~\cite{Ra82,Manton:2004tk,Vachaspati:2006zz}. Examples for these 
solutions are Skyrmions~\cite{Skyrme:1961vq,Zahed:1986qz}, monopoles in $(3+1)$ 
dimensions and vortices, strings and lumps in $(2+1)$ dimensions. There are numerous 
applications in various branches of physics: in cosmology~\cite{Vilenkin:2000jqa}, 
condensed matter physics~\cite{Schollwock:2004aa,Nagasoa:2013}, as well as 
hadron and nuclear physics~\cite{Weigel:2008zz}. We point to those textbooks 
and review articles for more details and further references.

The field equations minimize the classical energy. In particle physics applications
this energy represents the leading contribution to the particle masses. Since the early 
studies~\cite{Adkins:1983ya} of baryon properties in the Skyrme model it is known that 
these predictions exceed the actual masses by 30\% or more. It has therefore been 
conjectured that quantum corrections reduce these predictions appropriately~\cite{Meier:1996ng}.
Unfortunately the Skyrme model is not a renormalizable theory and these corrections cannot
be determined unambiguously. To nevertheless gain insight on whether such corrections can 
cause the conjectured effect, it is appropriate to investigate the vacuum polarization 
energy (VPE), which is the leading quantum correction to the classical 
soliton energy, in a renormalizable theory. Moreover, the Skyrme model with pion fields 
only has various deficiencies when it comes to the description of baryon properties. 
These deficiencies are overcome by adding massive vector mesons ($\rho$, 
$\omega$)~\cite{Meissner:1987ge,Schwesinger:1988af,Weigel:2008zz}, which are described 
by Proca fields that interact with the pion fields. This makes the exploration of the VPE
for interacting Proca fields in renormalizable field theory a very worthwhile topic.

The VPE is the renormalized shift of the zero point energies for the quantum
fluctuations about the soliton background. For a static background the bosonic
VPE is formally given by
\begin{equation}
E_{\rm VPE}=\frac{1}{2}\sum_k\left(\omega_k-\omega_k^{(0)}\right)+E_{\rm CT}\,.
\label{eq:vpeI}
\end{equation}
Here $\omega_k$ and $\omega_k^{(0)}$ are the energy eigenvalues of the fluctuations
about the background of the soliton and the translationally invariant vacuum,
respectively. Furthermore $E_{\rm CT}$ is the counterterm contribution that
implements the renormalization. It is part of the model definition but needs to
be adjusted at each order in the perturbation/loop expansion.

It has been corroborated that spectral methods, which
utilize scattering data for the quantum fluctuations about the static potential 
generated by the soliton are a very efficient technique to compute the 
VPE~\cite{Graham:2009zz}. This is especially the case for implementing 
standard renormalization conditions because the Born series of the scattering 
data is equivalent to the Feynman series of the one-loop effective action. 
Even more, the analytic properties of the Jost function, which is a particular 
solution to the scattering wave-equation, allow us to express the VPE as a 
single integral over imaginary momenta, and makes the method even more 
efficient~\cite{Graham:2022rqk}. For a particular model it is therefore 
compulsory to ensure that there is no obstacle for the underlying analytic 
continuation. Here we will discuss two situations for which such obstacles 
seem to exist. 

The VPE can also be computed from scattering formulations of the Green's
functions~\cite{Baacke:1989sb,Graham:2002xq}. Alternatively, the fluctuation
determinant is directly computed (or estimated) within heat kernel methods 
\cite{Bordag:1994jz} in conjuction with $\zeta$-function renormalization
\cite{Elizalde:1995hck}, or by the world line formalism \cite{Gies:2003cv}. Other 
approaches relate the quantum fields with and without the soliton background 
by a displacement operator \cite{Evslin:2019xte}, conduct derivative expansions
\cite{Aitchison:1985pp}, or apply the Gel'fand-Yaglom method \cite{Gelfand:1959nq};
just to name a few other techniques. The implementation
of standard renormalization conditions is not too obvious in most of these
approaches. This is even more the case when these techniques are applied
to models that contain quadratic divergences. Also, the unbiased reader should
be able to easily assess the superior efficiency of the spectral methods when
comparing the discussion in Sect.~\ref{sec:spectral} with, for example, the heat
kernel formalism detailed in the appendix of Ref.~\cite{AlonsoIzquierdo:2002eb}
or the lengthy and highly technical calculations in Ref.~\cite{Evslin:2019xte}.

After this introduction we will discuss two subtleties arising from (i) the
Born approximation in the presence of a mass gap and (ii) the quantization of
the Proca field. We will then briefly review the spectral methods approach to
the VPE in Chapter \ref{sec:spectral}. We will especially explain the effectiveness
of the imaginary momentum formulation. In Chapter \ref{sec:real} we will consider
a toy model for two scalar fields with different masses. We will present a solution
to the above mentioned Born obstacle and, by numerical simulation, we will verify that
the real and imaginary momentum formulations yield identical results. In Chapter
\ref{sec:Proca} we will introduce a Proca model in $D=1+1$ space-time dimensions and
construct its soliton solution. Thereafter, in Chapter \ref{sec:scatter}, we will
investigate the scattering problem in that model with emphasis on the role of the
non-standard normalization of the longitudinal mode. Chapter \ref{sec:num} contains
our numerical results for the VPE of the Proca soliton. We will briefly summarize
and outline related future projects in Chapter~\ref{sec:concl}.

\section{Particular subtleties}
\label{issues}
For the scattering problem for two (or more) fields with masses $m_1\le m_2$, the 
Born approximation to the sum of the phase shifts contains a contribution 
proportional to 
$$
\frac{1}{\sqrt{k^2-m_2^2+m_1^2}}\int dx\, v_{>}(x)\,,
$$
where $v_{>}$ is the self-interaction potential of the heavier particle. Obviously 
this is ill-defined (or imaginary) for momenta $k\in\left[0,\sqrt{m_2^2-m_1^2}\right]$ 
which correspond to energies in the mass gap. However, the Born approximation is a 
crucial element for implementing standard renormalization conditions. In 
Ref.~\cite{Weigel:2017kgy} this problem was circumvented by analytical continuation 
$k=\imu t$ so that the denominator becomes $\pm\imu\sqrt{t^2+m_2^2-m_1^2}$ and is 
only needed for $t\ge m_1$. Hence it is important to verify the validity of the analytic 
continuation when computing the VPE. It is needless to mention that the mass gap 
problem is also present in the above mentioned vector meson extensions of the 
Skyrme model as $\rho$ and $\omega$ are more than five times as heavy as the pion.

We face another possible obstacle in the Proca model for a vector meson field 
$V_\alpha$ with mass $\mu$. For a free Proca field in $D=1+1$ dimensions this field 
only has temporal ($V_0$) and longitudinal ($V_1$) components. The former is not 
dynamical (its time derivative does not appear in the Lagrangian) and its elimination 
yields an extended relation between the field velocity and the canonical momentum:
\begin{equation}
\Pi_1(t,x)=\dot{V}_1(t,x)+\frac{1}{\mu^2}\Pi^{\prime\prime}_1(t,x)\,.
\label{eq:canonical}\end{equation}
Here, and in what follows, dots and primes are time and space derivatives, respectively. 
The second term in Eq.~(\ref{eq:canonical}) requires the field decomposition
\begin{equation}
V_1(t,x)=\int \frac{dk}{2\pi(2\omega)}\,\frac{\omega}{\mu}
\left[a^\dagger(k){\rm e}^{-\imu(\omega t-kx)}+a(k){\rm e}^{\imu(\omega t-kx)}\right]\,,
\label{eq:Vdecomp}\end{equation}
so that $a(k)$ and $a^\dagger(k)$ are respectively annihilation and creation operators
for vector particles with energy $\omega=\sqrt{k^2+\mu^2}$. While the energy factor in 
the integration measure is standard\footnote{It is usually compensated by the very same 
factor in the commutation relation 
$\left[a(k),a^\dagger(k^\prime)\right]=2\pi(2\omega)\delta(k-k^\prime)$.},
its appearance in the ratio $\frac{\omega}{\mu}$ is unconventional and the corresponding 
square root discontinuity might hamper the analytic continuation in the momentum variable $k$.

\section{Brief review of spectral methods}
\label{sec:spectral}

The formal sum in Eq.~(\ref{eq:vpeI}) can be expressed as a discrete sum over 
bound states plus a continuum integral over scattering states. The latter are 
labeled by their momentum~$k$, with $\omega_k=\sqrt{k^2+m^2}$. That integral 
is conveniently evaluated using the Friedel-Krein formula~\cite{Faulkner:1977aa}, 
\begin{equation}
\Delta\rho=\frac{1}{\pi}\frac{d\delta(k)}{dk}
\label{eq:Krein}
\end{equation}
for the change in the density of states, $\Delta\rho$, generated by the soliton
background. Here we imply the sum over channels. For the projects mentioned in 
the introduction it suffices to consider the case of one space dimension. When the 
potential is reflection invariant this sum is over channels with spatially symmetric 
and anti-symmetric wave-functions and $\delta(k)$ is the sum of the respective 
phase shifts. When needed, we will refer to this sum as the total phase shift. 
We then evaluate the continuum part in Eq.~(\ref{eq:vpeI}) as an integral with 
the measure $\Delta\rho dk$
\begin{align}
E_{\rm VPE}&=\frac{1}{2}\sum_j^{\rm b.s.}\omega_j
+\int\frac{dk}{2\pi}\sqrt{k^2+m^2}\frac{d\delta(k)}{dk}+E_{\rm CT}\cr
&=\frac{1}{2}\sum^{\rm b.s.}_j\omega_j
+\int_0^\infty\frac{dk}{2\pi}\sqrt{k^2+m^2}\frac{d}{dk}\left[\delta(k)-\delta_B(k)\right]
+E_{\rm FD}+E_{\rm CT}\cr
&=\frac{1}{2}\sum^{\rm b.s.}_j(\omega_j-m)
+\int_0^\infty\frac{dk}{2\pi}\left[\sqrt{k^2+m^2}-m\right]
\frac{d}{dk}\left[\delta(k)-\delta_B(k)\right]
+E_{\rm FD}+E_{\rm CT}\,.
\label{eq:vpeII}\end{align}
The discrete sum now only involves the isolated bound state energies $|\omega_j|<m$.
In the second line we have subtracted the leading terms of the Born series for the
phase shift. The (smallest) number of these terms is determined such that the momentum
integral is finite. The Born contributions to the VPE can alternatively be expressed 
as Feynman diagrams, $E_{\rm FD}$, which we use to add back the preceding subtractions 
under the integral. These diagrams are regularized by standard means ({\it e.g.} 
dimensional regularization) and $E_{\rm FD}+E_{\rm CT}$ remains finite when the 
regulator is removed. For boson models in one space dimension the only required 
Feynman diagram is proportional to the spatial integral over the background potential. 
Thus we can implement the no-tadpole renormalization condition in the form 
$E_{\rm FD}+E_{\rm CT}=0$. In the third line of Eq.~(\ref{eq:vpeII}) we have 
used Levinson's theorem that relates the phase shift at zero momentum to the
number of bound states\footnote{See Ref.~\cite{Barton} for the formulation of 
Levinson's theorem in one space dimension.}. 

A central element of potential scattering theory is the Jost solution $f_k(x)$. It solves 
the wave-equation subject to the asymptotic condition
\begin{equation}
\lim_{x\to\infty}f_k(x)\,{\rm e}^{-\imu kx}=1
\label{eq:asymp}\end{equation}
and is analytic in the upper half momentum plane, {\it i.e.} for 
${\sf Im}(k)\ge0$ \cite{Newton:1982qc}. For spatially symmetric potentials
the Jost function $F(k)$ is extracted from $f_k(0)$ and $f^\prime_k(0)$. (Below we will 
give more details when discussing particular models.) Obviously the Jost function is also
analytic for ${\sf Im}(k)\ge0$ and for real $k$ its phase is the scattering phase shift 
$$
F(k)=|F(k)|{\rm e}^{-\imu\delta(k)}\,.
$$
While the modulus $|F(k)|$ is an even function of real $k$, the phase is odd. This
can be easily understood from the asymptotic condition above: Since the wave-equation is
real, that condition implies that $k\to-k$ corresponds to complex conjugation. For an 
arbitrary function $g=g(k^2)$ we thus have
$$
\int_0^\infty\frac{dk}{2\pi}g(k^2)\frac{d}{dk}\left[\delta(k)\right]_B
=\frac{\imu}{2}\int_{-\infty}^\infty\frac{dk}{2\pi}g(k^2)\frac{d}{dk}\left[\ln F(k)\right]_B\,,
$$
where the subscript indicates the necessary subtractions from the Born series. We use
the Jost function to compute Eq.~(\ref{eq:vpeII}) by completing the contour in the upper 
half $k$-plane. The Born subtraction ensures that there is no contribution from the 
semi-circle at $|k|\to\infty$. It remains to collect the residues and bypass the 
branch-cut along the imaginary axis induced by $g(k^2)=\sqrt{k^2+m^2}-m$. Another 
well-established property of the Jost function is that it has single roots which are 
located on the imaginary axis at the wave-numbers of the bound state energies \cite{Newton:1982qc}: 
$\kappa_j=\sqrt{m^2-\omega_j^2}$ which implies 
$$
\frac{d}{dk}\ln F(k)\approx \frac{1}{k-\imu\kappa_j}
\qquad {\rm for}\qquad k\approx \imu\kappa_j\,.
$$
Hence the pole contributions cancel against the (explicit) bound state sum in Eq.~(\ref{eq:vpeII}).
The branch cut is along the imaginary axis starting at $k=\imu m$.  We introduce $k=\imu t$, 
recognize that the square root discontinuity is $2\imu\sqrt{t^2-m^2}$ and finally obtain
\begin{equation}
E_{\rm VPE}=\int_m^\infty\frac{dt}{2\pi}\frac{t}{\sqrt{t^2-m^2}}\,\left[\ln F(\imu t)\right]_B
\label{eq:master}\end{equation}
for a boson theory in $D=1+1$ dimensions within the no-tadpole renormalization scheme.
In the renowned sine-Gordon and $\phi^4$-kink models the total Jost functions are
known~\cite{Graham:2022rqk}
$$
F_{\rm sG}(\imu t)=\frac{t-m}{t+m}
\qquad {\rm and}\qquad 
F_{\rm kink}(\imu t)=\frac{t-m}{t+m}\frac{2t-m}{2t+m}\,.
$$
Direct integration yields the well-established results for the VPE:
$-\frac{m}{\pi}$ and $\frac{m}{12\pi}\left(\sqrt{3}\pi-18\right)$, 
respectively \cite{Ra82}. This calculation clearly illustrates the effectiveness
of the imaginary axis formulation.

\section{Real axis calculation}
\label{sec:real}
As we will see, the effectiveness of the imaginary axis formulation is 
even more pronounced for systems with coupled scattering channels and 
different mass parameters. We will always consider two such channels 
with the convention that the mass parameters are ordered $m_1\le m_2$.
The other case is merely a matter of re-labeling.

The imaginary axis formulation has already been established some time
ago \cite{Weigel:2017kgy}. Nevertheless and precisely because of that,
it is necessary to verify that it agrees with the real axis approach,
Eq.~(\ref{eq:vpeI}). The latter is hampered not only by the Born obstacle
mentioned in the introduction but also by the proper identification of the 
phase shift, which in the direct numerical simulation is obtained in the 
interval $\left[-\pi,\pi\right]$. As will be discussed below, with the 
existence of threshold cusps that endeavor may be cumbersome. We will 
investigate these issues within a toy model defined by a spatially symmetric 
$2\times2$ potential matrix $V(x)=(v_{ij}(x))$. We combine the two fields 
$\phi_1$ and $\phi_2$ into a two-component array 
$\psi=\left(\phi_1,\phi_2\right)^{\rm t}$ which obeys the wave-equation
\begin{equation}
\ddot{\psi}(x,t)-\psi^{\prime\prime}(x,t)
=-M^2 \psi(x,t)-V(x)\psi(x,t)\,.
\label{eq:toy1}\end{equation}
with the diagonal mass matrix $M^2={\rm diag}\left(m_1^2,m_2^2\right)$.
Since the potential is static we can factorize the time dependence as
$\psi(x,t)=\psi_k(x){\rm e}^{-\imu\omega t}$ where $k=\sqrt{\omega^2-m_1^2}$ is 
a unique label. Furthermore we define a matrix of Jost solutions where the
entries of a particular column are the two fields $\phi_1$ and $\phi_2$ while
the different columns refer to solutions to the wave-equation that 
asymptotically have outgoing plane waves for only one of the two fields. 
To be precise we write
\begin{equation}
F_k(x)=Z_k(x)\begin{pmatrix} {\rm e}^{\imu kx} & 0 \cr 0 & {\rm e}^{\imu k_2x}\end{pmatrix}
\qquad {\rm with}\qquad
k_2=k_2(k)\equiv k\sqrt{1-\frac{m_2^2-m_1^2}{\left[k+\imu0^{+}\right]^2}}\,.
\label{eq:JostMatrix1} \end{equation}
The particular form of the dependent momentum $k_2$ ensures the pertinent 
behavior under $k\to-k$ (symmetric in the gap, but anti-symmetric outside) 
and that all additional singularities will reside in the lower half complex 
$k$-plane \cite{Weigel:2017kgy}. The coefficient matrix is subject to the 
second order differential equation
\begin{equation}
Z^{\prime\prime}_k(x)=-2Z^\prime_k(x)D(k)+\left[M^2,Z_k(x)\right]+V(x)Z_k(x)
\qquad {\rm with}\qquad
D(k)=\imu\begin{pmatrix}k & 0 \cr 0 & k_2\end{pmatrix}\,.
\label{eq:JostDEQ}\end{equation}
The asymptotic condition, Eq.~(\ref{eq:asymp}) translates into
$\lim_{x\to\infty}Z_k(x)=\ID$. The Born series is straightforwardly constructed
by iterating $Z_k(x)=\ID+Z^{(1)}_k(x)+Z^{(2)}_k(x)+\ldots$, where the superscripts
refer to the order in the background potential $V(x)$. That is, $V(x)$ is the source for
$Z^{(1)}_k(x)$, $V(x)Z^{(1)}_k(x)$ is the source for $Z^{(2)}_k(x)$ and so on.
All $Z^{(l)}_k(x)$ vanish at spatial infinity.

The scattering wave-functions are linear combinations of $F_k(x)$ and
$F^\ast_k(x)=F_{-k}(x)$ and the relative weight is the scattering matrix.
In the anti-symmetric ($-$) and symmetric~($+$) channels the wave-function
and its derivative, respectively, vanish at the center of the potential. 
From that we get the scattering matrices. Since we are only interested 
in the total phase shift we consider
${\rm det}\left[S_{\pm}(k)\right]={\rm det}\left[F_\pm(-k)F^{-1}_\pm(k)\right]$
with the Jost matrices
\begin{equation}
F_{+}(k)=\lim_{x\to0}\left[Z^\prime(k,x)D^{-1}(k)+Z(k,x)\right]
\qquad {\rm and}\qquad
F_{-}(k)=\lim_{x\to0}Z(k,x)\,.
\label{eq:defjostsym}\end{equation}
Finally, the total phase shift is given by
\begin{equation}
\delta(k)=-{\sf Im}\left({\rm ln}{\rm det}\left[F_{+}(k)F_{-}(k)\right]\right)\,.
\label{eq:phaseshift}\end{equation}
One reason for the $\imu0^{+}$ prescription in Eq.~(\ref{eq:JostMatrix1}) is to ensure 
that ${\rm e}^{\imu k_2x}$ decays exponentially for momenta in the mass gap regardless 
of the sign of $k$. For that reason we can use Eq.~(\ref{eq:phaseshift}) also in that 
momentum regime when the second particle channel is closed.

The fact that the wave-function in the symmetric channel has a boundary condition 
on its derivative leads to a slight modification of Levinson's theorem in one space
dimension~\cite{Barton}:
\begin{equation}
\delta(0)=\pi\left(n-\frac{1}{2}\right)\,.
\label{eq:Levinson}\end{equation}
As $\delta(k)$ is the sum of the phase shifts in the symmetric and anti-symmetric
channels, so is $n$ the total number of bound states from these two channels.

For later consideration the Born approximation is extracted from
\begin{align*}
\lim_{x\to0}
&{\rm tr}{\rm ln}\left(\left[Z^{(1)\prime}(k,x)D^{-1}(k)+\ID+Z^{(1)}(k,x)\right]
\left[\ID+Z^{(1)}(k,x)\right]\right)
\\ &\hspace{2cm}\approx \lim_{x\to0}
{\rm tr}\left(Z^{(1)\prime}(k,x)D^{-1}(k)+2Z^{(1)}(k,x)\right)\,.
\end{align*}
Multiplying the first order expansion of equation~(\ref{eq:JostDEQ}) with $D^{-1}(k)$ 
from the right and integrating from zero to infinity yields\footnote{Note that
${\rm tr}\left(\left[M^2,Z^{(1)}(k,x)\right]D^{-1}(k)\right)=0$.}
\begin{equation}
\delta_B(k)=-\frac{1}{k}\int_0^\infty dx\, v_{11}(x)
-\frac{1}{k_2}\int_0^\infty dx\, v_{22}(x)\,.
\label{eq:Born}\end{equation}
Of course, this expression is only well-defined outside the mass gap where
$k_2\in\mathbb{R}$.

\begin{figure}
\centerline{
\includegraphics[width=6.5cm,height=4cm]{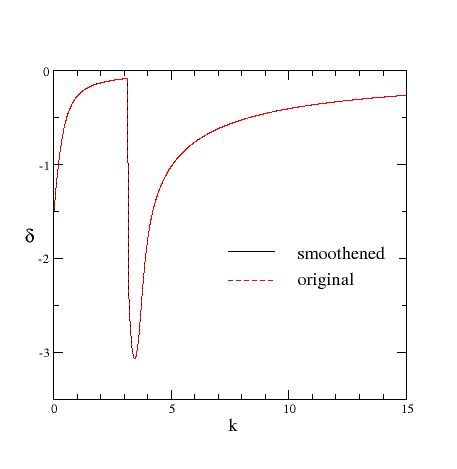}\hspace{5mm}
\includegraphics[width=6.5cm,height=4cm]{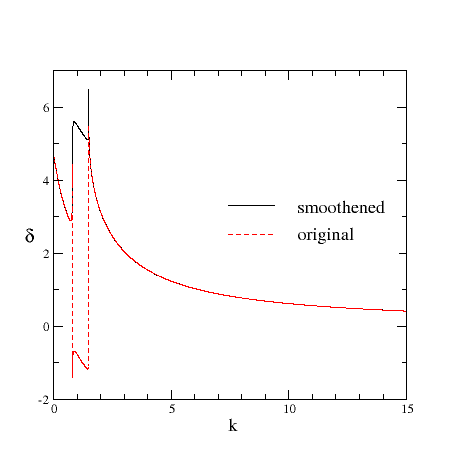}}
\caption{\label{fig:toy1}Total phase shift in the toy model for repulsive
(left) and attractive (right) potentials. The parameters for the repulsive 
case are $m_1=1.5$, $m_2=3.5$, ${\overline v}_{11}=4.0, {\overline v}_{22}=2.5$ 
and ${\overline v}_{12}={\overline v}_{21}=2.0$. In that case the two lines are 
on top of each other. The attractive potential is parameterized by $m_1=2.0$, 
$m_2=2.5$, ${\overline v}_{11}=-4.0, {\overline v}_{22}=-3.0$ 
and ${\overline v}_{12}={\overline v}_{21}=-0.5$.}
\end{figure}

In numerical simulations equation~(\ref{eq:phaseshift}) only produces results
in the interval $\left[-\pi,\pi\right]$ which, in particular when there are
bound states, does not produce a smooth function of~$k$ that vanishes as 
$k\to\infty$. To this end we apply a smoothening algorithm by evaluating
equation~(\ref{eq:phaseshift}) on a dense grid for discretized momenta and 
add multiples of $2\pi$ on identified intervals such that the difference of 
the phase between two neighboring momenta does not exceed $1.05\pi$. This 
produces a smooth function to which we eventually add $\pm2\pi$ to accommodate
the large $k$ limit. However, this procedure is delicate when the actual phase 
shift indeed has sharp cusps as it is the case at the top end of the mass gap 
or for Feshbach resonances \cite{Feshbach:1958nx} just below the mass gap.
It may therefore still happen that this smoothening procedure fails to yield
the correct phase shift within the gap. We can test for that 
by inverting Levinson's theorem, Eq.~(\ref{eq:Levinson}): We determine 
the bound state energies numerically by applying a shooting method to the 
differential equation for $\psi_{\imu\kappa}(x)$ such that the wave-function 
is continuous and vanishes exponentially as $x\to\infty$. This is possible
only for discrete values of $|\omega_j|\le m_1$. Once we have found these
bound state energies, we also know $n$ and thus the correct value $\delta(0)$.
If it does not match the result from the smoothing method, we need to add 
(or subtract) appropriate integer multiples of $2\pi(m_1-m_2)/2\pi=m_1-m_2$
to the VPE. Eventually we will also probe the so computed VPE against the imaginary 
axis calculation. In Fig.~\ref{fig:toy1} we display characteristic results for this
smoothening procedure for the toy model potential 
$v_{ij}(x)={\overline v}_{ij}{\rm e}^{-x^2}$. 
For attractive potentials there are no bound states and the original and 
smoothened phases are identical. On the other hand, for (strongly)
attractive potentials this need not be the case. For the example with the
attractive potential used in Fig.~\ref{fig:toy1} we find bound states at 
$\omega_1=1.256$ and $\omega_2=1.960$ in the symmetric and anti-symmetric 
channels, respectively. Hence we should have $\delta(0)=\frac{3}{2}\pi$, as 
is obeyed by the smoothened phase. Yet, the violent behavior of the original 
phase shift indeed indicates that there are numerical subtleties.

We next address the problem associated with the Born subtraction. Obviously we cannot 
use the exact Born approximation, Eq.~(\ref{eq:Born}) in Eq.~(\ref{eq:vpeII}).
Rather, we consider
\begin{equation}
\widetilde{E}_{\rm VPE}=\frac{1}{2}\sum_k^{\rm b.s.}(\omega_k-m_1)
-\int_0^\infty\frac{dk}{2\pi}\frac{k}{\sqrt{k^2+m_1^2}}
\left[\delta(k)-\Delta(k)\right]\,,
\label{eq:tildeVPE}\end{equation}
with $\Delta_B(k)=-\frac{1}{k}\int_0^\infty dx \left[v_{11}(x)+v_{22}(x)\right]$.
We will show that the difference to $E_{\rm VPE}$ can be calculated in two
ways that yield identical results: (i) by a reparameterization of the momentum
integral and (ii) the difference of two Feynman diagrams.

Since the Born approximation only concerns the diagonal elements of the 
potential matrix it suffices to consider a single channel problem with a spatially 
symmetric potential $v(x)$ in the channel with the heavier mass. With the 
definition $\langle V\rangle=-\int_0^\infty dx\,v(x)$ the corresponding VPE is
\begin{align*}
E_2&=\frac{1}{2}\sum_j^{\rm b.s.}\left(\omega_j-m_2\right)
-\int\limits_0^\infty\frac{dk}{2\pi}\frac{k}{\sqrt{k^2+m_2^2}}
\left[\delta(k)-\frac{\langle V\rangle}{k}\right]\\
&=\frac{1}{2}\sum_j^{\rm b.s.}\left(\omega_j-m_2\right)
-\int\limits_{\sqrt{m_2^2-m_1^2}}^\infty\frac{dq}{2\pi}\frac{q}{\sqrt{q^2+m_1^2}}
\left[\delta\left(\sqrt{q^2-m_2^2+m_1^2}\right)
-\frac{\langle V\rangle}{\sqrt{q^2-m_2^2+m_1^2}}\right]\,,
\end{align*}
with the new integration variable $q=\sqrt{k^2-m_1^2+m_2^2}$. According to
equation~(\ref{eq:tildeVPE}) we would calculate
$$
\widetilde{E}_2=\frac{1}{2}\sum_j^{\rm b.s.}\left(\omega_j-m_1\right)
-\int\limits_{0}^\infty\frac{dq}{2\pi}\frac{q}{\sqrt{q^2+m_1^2}}
\left[F\left(q\right)-\frac{\langle V\rangle}{q}\right]\,,
$$
with
$$
F(q)=\begin{cases}
n_2\pi=\delta(0) & {\rm for}\quad q^2\le m_2^2-m_1^2\\
\delta\left(\sqrt{q^2-m_2^2+m_1^2}\right) & {\rm for}\quad q^2\ge m_2^2-m_1^2\,,
\end{cases}
$$
where $n_2$ is the number of bound states in that channel and the smoothening
produces the $q^2\le m_2^2-m_1^2$ part. The contribution of that part is obtained from
$$
\int\limits_0^{\sqrt{m_2^2-m_1^2}}\frac{dq}{2\pi}\frac{n_2\pi q}{\sqrt{q^2+m_1^2}}
=\frac{n_2}{2}\sqrt{q^2+m_1^2}\Big|_0^{\sqrt{m_2^2-m_1^2}}
=\frac{1}{2}\sum_j^{\rm b.s.}\left(m_2-m_1\right)\,.
$$
This allows us to write
\begin{align*}
\widetilde{E}_2&=\frac{1}{2}\sum_j^{\rm b.s.}\left(\omega_j-m_2\right)
+\hspace{-2mm}\int\limits_0^{\sqrt{m_2^2-m_1^2}}\frac{dq}{2\pi}
\frac{\langle V\rangle}{\sqrt{q^2+m_1^2}}
\\ & \hspace{3cm}
-\int\limits_{\sqrt{m_2^2-m_1^2}}^{\infty}\frac{dq}{2\pi}\frac{q}{\sqrt{q^2+m_1^2}}
\left[\delta\left(\sqrt{q^2-m_2^2+m_1^2}\right)-\frac{\langle V\rangle}{q}\right]\,,
\end{align*}
resulting in
\begin{eqnarray}
E_2-\widetilde{E}_2
&=&\hspace{-2mm}-\hspace{-3mm}
\int\limits_0^{\sqrt{m_2^2-m_1^2}}\frac{dq}{2\pi}\frac{\langle V\rangle}{\sqrt{q^2+m_1^2}}
+\int\limits_{\sqrt{m_2^2-m_1^2}}^{\infty}\frac{dq}{2\pi}\frac{q}{\sqrt{q^2+m_1^2}}
\left[\frac{\langle V\rangle}{\sqrt{q^2-m_2^2+m_1^2}}-\frac{\langle V\rangle}{q}\right]\cr
&=&\int\limits_0^\infty \frac{dq}{2\pi} \left[\frac{\langle V\rangle}{\sqrt{q^2+m_2^2}}
-\frac{\langle V\rangle}{\sqrt{q^2+m_1^2}}\right]
=\frac{\langle V\rangle}{2\pi}\ln\frac{m_1}{m_2}\,.
\label{eq:DiffBorn0}\end{eqnarray}
Next we look at the one-loop effective action arising from the fluctuations and single out the 
$\mathcal{O}(V)$ contribution:
\begin{align*}
\Delta\mathcal{A}^{(1)}&=
\frac{\imu}{2}{\rm Tr}\left[\left(\partial^2+m_1^2-\imu\epsilon\right)^{-1}v
-\left(\partial^2+m_2^2-\imu\epsilon\right)^{-1}v\right]\cr
&=\frac{\imu}{2}\int \frac{d^2k}{(2\pi)^2}\left[
\left(-k^2+m_1^2-\imu\epsilon\right)^{-1}-\left(-k^2+m_2^2-\imu\epsilon\right)^{-1}\right]
\widetilde{v}(0)\,,
\end{align*}
where $\widetilde{v}$ is the Fourier transform of $v(x)$ so that
$\widetilde{v}(0)=\int d^2x\,v(x)=-2\langle V\rangle T$, with $T$ denoting an (infinitely large) 
time interval. The above integral is straightforwardly evaluated by Wick rotation yielding
\begin{equation}
\Delta\mathcal{A}^{(1)}=-\frac{\langle V\rangle T}{2\pi}\ln\frac{m_1}{m_2}\,.
\label{eq:DiffAction}\end{equation}
The corresponding effective energy matches Eq.~(\ref{eq:DiffBorn0}). Hence using Eq.~(\ref{eq:tildeVPE}) 
for the real axis calculation and correcting it with Eq.~(\ref{eq:DiffBorn0}) properly implements the 
no-tadpole condition. Essentially we regulate the ultra-violet divergence from the heavier particle 
by a Feynman diagram with the lighter particle in the loop. We may consider this treatment as a 
variant of the Pauli-Villars regularization scheme. In the two channel problem we write the
correction as 
\begin{equation}
E_{\rm vac}-\widetilde{E}_{\rm vac}=\frac{\langle V\rangle}{2\pi}\ln\frac{m_{<}}{m_{>}}
\qquad{\rm with}\qquad 
\langle V\rangle=-\int_0^\infty dx\, v_{>}(x)\,.
\label{eq:DiffBorn}\end{equation}
Here the subscripts refer to the smaller ($<$) and larger ($>$) of the two masses and
$v_{>}$ again is the self-interaction component of potential matrix for the heavier particle.

The goal is to compare the numerical results from the approach outlined above to the imaginary 
momentum procedure discussed in Chap.~\ref{sec:spectral}. The latter has been generalized to the
two-component system in Ref.~\cite{Weigel:2017kgy}. We need to solve Eq.~(\ref{eq:JostDEQ}) 
with the replacements $k\to t=\imu k$ and $k_2\to \imu\sqrt{t^2+m_2^2-m_1^2}$ to write the 
VPE as
\begin{equation}
E_{\rm vac} \equiv 
\frac{1}{2\pi}\int_{0}^{\infty} d\tau\,
\left[\nu(t)-\nu^{(1)}(t)\right]_{t=\sqrt{\tau^2+m_1^2}}\,,
\label{eq:Evac} 
\end{equation}
with
\begin{equation}
\nu(t) \equiv {\rm ln}\,{\rm det}\left[F_{+}(\imu t) \,F_{-}(\imu t)\right]
\quad {\rm and}\quad
\nu^{(1)}(t)=\int_{0}^\infty dx\, \left[\frac{v_{11}(x)}{t}
+\frac{v_{22}(x)}{\sqrt{t^2+m_2^2-m_1^2}}\right]\,.
\label{eq:defnu}
\end{equation}
Clearly there is no singularity in the Born approximation, $\nu^{(1)}$. We also 
note that the $F_{\pm}(\imu t)$ are real-valued matrices. In Tab.~\ref{tab:toy} we compare 
the results for the two scenarios of Fig.~\ref{fig:toy1}.
\begin{table}
\centerline{
\begin{tabular}{l|lll|l}
& $\widetilde{E}_{\rm vac}$
& $E_{\rm vac}-\widetilde{E}_{\rm vac}$ & $E_{\rm vac}$ & $E_{\rm vac}$\\
& {\small Eq.~(\ref{eq:tildeVPE})} &  {\small Eq.~(\ref{eq:DiffBorn})} 
&& {\small Eq.~(\ref{eq:Evac})}\\
\hline
repulsive  & -0.5102 & 0.4780 & -0.0322 & -0.0324 \\
attractive & -0.0872 & -0.0944 & -0.1817 & -0.1821 \\[2mm]
\end{tabular}}
\caption{\label{tab:toy}Comparison of real and imaginary axis calculation of the VPE
for the Gaussian potential matrix $v_{ij}={\overline v}_{ij}{\rm e}^{-x^2}$. Model 
parameters are as in Fig.~\ref{fig:toy1}.}
\end{table}
The agreement of the two approaches could not be clearer. We have performed numerous
such comparisons \cite{Petersen:2024} and never obtained mismatches in the leading three
significant digits after rounding the fourth one.

The considerable lower cost of computing time makes the imaginary axis approach
significantly more efficient. This is mainly caused by the smoothing procedure
which requires a dense discretization for the real momenta. Also the fact
that the real axis approach solves a differential equation for a complex matrix
rather than a real one adds computing time. On top there is the advantage
that the imaginary axis approach does not require to explicitly find the bound 
state energies. Nevertheless this toy model exercise impressively confirms the 
equivalence of the real and imaginary momentum computations of the VPE, even in 
the presence of potential branch cuts arising from energy thresholds. We will 
use that knowledge to explore the potential non-analyticity in a vector meson 
Proca model in Chap. \ref{sec:scatter}. But first we need to construct that soliton.

\section{Proca soliton}
\label{sec:Proca}

\begin{figure}
\centerline{
\includegraphics[width=4cm,height=4cm]{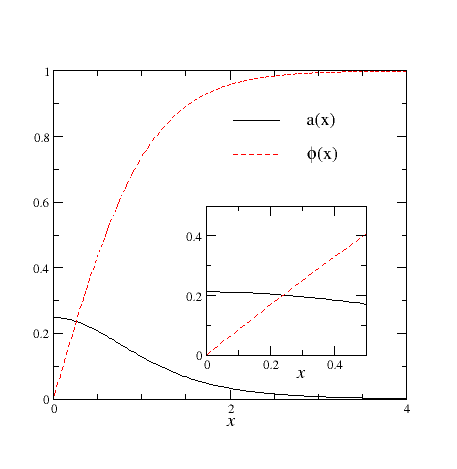}\hspace{0.5cm}
\includegraphics[width=4cm,height=4cm]{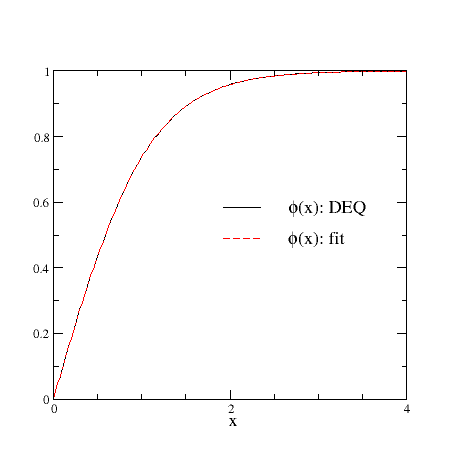}\hspace{0.5cm}
\includegraphics[width=4cm,height=4cm]{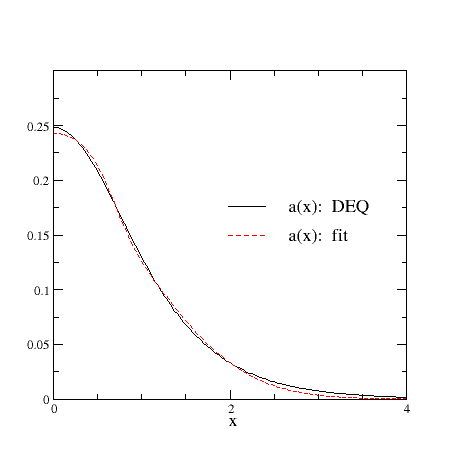}}
\caption{\label{fig:sol}Soliton profiles for $g=1.0$ and $\mu=1.5$ (left panel).
The inlay corroborates that $a^\prime(0)=0$.
Also shown are the fitted profiles for $\phi$ (middle panel) and $a$ (right panel)
in comparison with the solutions to Eq.~(\ref{eq:eom}); labeled 'DEQ'.}
\end{figure}

We consider a Lagrangian in $D=1+1$ space time dimensions with two real fields: 
a scalar ($\Phi$) and a massive vector meson ($V_\alpha$)
\begin{equation}
\mathcal{L}=\frac{1}{2}\partial_\alpha \Phi \partial^\alpha\Phi
-\frac{1}{4}\left(\partial_\alpha V_\beta-\partial_\beta V_\alpha\right)^2
+\frac{\mu^2}{2}V_\alpha V^\alpha -\frac{1}{2}\left(\Phi^2-1\right)^2
-g\left(1-\Phi^2\right)\epsilon^{\alpha\beta}V_\alpha\partial_\beta \Phi\,.
\label{eq:lag1}\end{equation}
The scale is set by the scalar meson mass $m_\phi=2$, the scalar self-interaction is that
of the $\phi^4$ kink model and the vector meson mass is $\mu$. The coupling is constructed 
such that it is at least cubic in the fluctuations about the possible vacuum configurations
($\Phi_0=\pm1$ and $V_0^\alpha=0$) and the $\epsilon$ tensor ensures that the field equations 
are consistent with $\partial^\alpha V_\alpha=0$. The vector meson mass $\mu$ and the coupling 
constant $g$ are the only tuneable model parameters. We have scaled variables, parameters
and fields to dimensionless quantities. This produced an overall factor for $\mathcal{L}$
which we do not explicitly write because it would only matter if we compared classical and 
quantum energies quantitatively. The model can be considered as the one-dimensional
reduction of the Skyrme model with an $\omega$ meson \cite{Adkins:1983nw}. This also 
motivates the soliton ansatz with a profile function only for the time component of
the vector meson field, $V^\alpha=(a(x),0)$ and $\Phi=\phi(x)$. The classical energy 
functional becomes
\begin{equation}
E_{\rm cl}=\frac{1}{2}\int dx \left\{\phi^{\prime2}+\left(\phi^2-1\right)^2
-a^{\prime2}-\mu^2a^2+2g(1-\phi^2)a\phi^\prime\right\}\,.
\label{eq:ecl}\end{equation} 
The variational principle yields the static equations
\begin{equation}
a^{\prime\prime}=\mu^2a-g\left(1-\phi^2\right)\phi^\prime
\qquad {\rm and}\qquad 
\phi^{\prime\prime}=2\phi\left(\phi^2-1\right)-g\left(1-\phi^2\right)a^\prime\,.
\label{eq:eom}\end{equation}
For $g=0$ they are solved by the ordinary kink, $\phi_{\rm K}={\rm tanh}(x)$ and $a\equiv0$.
Stable soliton solutions for $g\ne 0$ should also connect the two possible vacua 
$\phi_0=\pm1$ at positive and negative spatial infinity. This implies that $\phi$ is
odd under the reflection around its center $x_0$ with $\phi(x_0)=0$. We choose $x_0=0$
and find that $a(-x)=a(x)$. We solve Eqs.~(\ref{eq:eom}) with a shooting method
on the positive half-line, $x\ge0$ subject to the boundary conditions
\begin{equation}
\phi(0)=0\,,\quad a^\prime(0)=0
\qquad {\rm and}\qquad
\lim_{x\to\infty}\phi(x)=1\,,\quad
\lim_{x\to\infty}a(x)=0\,.
\label{eq:boundary}\end{equation}
The profiles on the negative half-line, $x\le0$ can be constructed via the above discussed 
reflection properties. A typical solution is shown in the left panel of Fig.~\ref{fig:sol}.
For all cases considered, we have verified that 
\begin{flalign}
&&\int_0^\infty dx \left\{\phi^{\prime2}-a^{\prime2}\right\}
=\int_0^\infty dx \left\{\left(\phi^2-1\right)^2-\mu^2a^2\right\} \hspace{-1cm}&\cr
&{\rm and}\qquad 
&\int_0^\infty dx \left\{a^{\prime2}+\mu^2a^2-g(1-\phi^2)a\phi^\prime\right\}=0 &\hspace{8cm}
\label{eq:scaling}\end{flalign}
are fulfilled. The first equation reflects Derrick's theorem \cite{Derrick:1964ww}
while the second is a consequence of stability under scaling the vector meson profile.
Numerically these profiles are only known at prescribed values of the coordinate $x$.
However, later in the scattering problem we will apply an adaptive step size algorithm
which requires the profiles at other $x$ values as well. Rather than implementing a
(CPU time costly) interpolation, we choose to fit the profiles to analytic functions.
A good choice is 
\begin{equation}
\phi_{\rm fit}(x)=a_0\tanh(a_1x)+a_2\tanh(a_3x)
\qquad{\rm and}\qquad
a_{\rm fit}(x)=b_0{\rm e}^{-b_1x^2}+b_0{\rm e}^{-b_1x^4}\,.
\label{eq:fit}\end{equation}
Even though there is some arbitrariness\footnote{The fit algorithm consistently produced
$a_0+a_2\approx1$, see also Eq.~(\ref{eq:fit2}) below.} in $\phi_{\rm fit}(x)$ we find 
fitting parameters that perfectly match the solution from the differential equation, as
seen in the middle panel of Fig.~\ref{fig:sol}. The fit to the vector profile shows 
some minor deviations from the actual solution, in particular asymptotically as may be
observed from the right panel in the same figure. In later applications we 
actually fit $a^\prime(x)$ directly. In any event, we are typically able to 
construct fits that violate the identities in Eq.~(\ref{eq:scaling}) by only about
one in a thousand or less. This can also seen from the data for the classical mass
in Tab.~\ref{tab:ecl}.
\begin{table}
\centerline{
\begin{tabular}{l|l|lllll}
$\mu$ & $g$ & 0.4 & 0.8 & 1.2 & 1.6 & 2.0\\
\hline
1.5 & $E_{\rm cl}$ & 1.356 & 1.422 & 1.532 & 1.679 & 1.858\\
    & $E_{\rm fit}$ & 1.356 & 1.423 & 1.533 & 1.680 & 1.860\\
\hline
2.5 & $E_{\rm cl}$ & 1.343 & 1.372 & 1.419 & 1.484 & 1.564\\
    & $E_{\rm fit}$ & 1.343 & 1.372 & 1.419 & 1.484 & 1.566\\[2mm]
\end{tabular}}
\caption{\label{tab:ecl}The classical energy, Eq.~(\ref{eq:ecl}) for the solutions
to the field equations,~(\ref{eq:eom}) and the fitted profiles, Eq.~(\ref{eq:fit}) as
functions of the coupling constant $g$ and two values of the vector meson mass~$m$.}
\end{table}
From that table we see that the classical energy increases with the coupling constant. 
A bit more surprising is that it decreases as the vector meson mass gets larger. We may
explain this by noting that for large $\mu$ the derivative term $a^{\prime2}$ may be
omitted against $\mu^2a^2$ and the field equation may be locally approximated by 
$$
a\approx \frac{g}{\mu^2}\left(1-\phi^2\right)\phi^\prime\,.
$$ 
Then the vector meson profile is no longer dynamical and we may approximate
the energy functional by
$$
E_{\rm cl}\approx\frac{1}{2}\int dx \left\{\phi^{\prime2}+\left(\phi^2-1\right)^2
+\frac{g^2}{\mu^2}(1-\phi^2)^2\phi^{\prime2}\right\}\,,
$$
indicating that with growing $\mu$ we are left with the pure kink model which has
classical energy $E^{\rm (K)}_{\rm cl}=\frac{4}{3}$ for the units used here.

\section{Scattering problem in Proca model}
\label{sec:scatter}
We formulate the scattering problem by introducing small amplitude fluctuations
about the above constructed soliton:
\begin{equation}
\Phi(x,t)=\phi(x)+{\rm e}^{-\imu \omega t}\eta(x)\,,
\quad
V_0(x,t)=a(x)+{\rm e}^{-\imu \omega t} u_0(x)
\quad{\rm and}\quad
V_1(x,t)={\rm e}^{-\imu \omega t} u_1(x)\,.
\label{eq:lin1} \end{equation}
The time dependences factorizes because the soliton is static and we omit
to explicitly write the frequency ($\omega$) dependence of the fluctuations
$\eta$, $u_0$ and $u_1$. With this parameterization the continuity equation 
$\partial^\alpha V_\alpha=0$ reads $u_0=-\frac{\imu}{\omega}\,u_1^\prime$. This 
allows us to eliminate $u_0$ from the linearized field equations and obtain
\begin{align}
u_1^{\prime\prime}&=\left(\mu^2-\omega^2\right)u_1
-\imu g\omega\left(1-\phi^2\right)\eta\\
\eta^{\prime\prime}&=\left(4-\omega^2\right)\eta+6\left(\phi^2-1\right)\eta
+2g\phi a^\prime\eta+\frac{\imu}{\omega}\mu^2\left(1-\phi^2\right)u_1
+g^2\left(1-\phi^2\right)^2\eta\,.
\label{eq:lin2}\end{align}
We immediately observe that the scattering problem is non-Hermitian. Rather, the 
coefficient functions of $\eta$ in the differential equation for $u_1$ and 
its counterpart in the differential equation for $\eta$ differ by factors
$\frac{\omega}{\mu}$. When we introduce the scaled vector fluctuation 
$\overline{u}_1$ via\footnote{See also Chap.~6 in Ref.~\cite{Schwesinger:1988af}.}
\begin{equation}
u_1=-\imu\frac{\omega}{\mu}\,\overline{u}_1\,,
\label{eq:lin3} \end{equation}
the fluctuation equations indeed assume an Hermitian form,
\begin{align}
\overline{u}_1^{\prime\prime}&=\left(\mu^2-\omega^2\right)\overline{u}_1
+g\mu\left(1-\phi^2\right)\eta\cr
\eta^{\prime\prime}&=\left(4-\omega^2\right)\eta
+\left[6\left(\phi^2-1\right)+2g\phi a^\prime+g^2\left(\phi^2-1\right)^2\right]\eta
+g\mu\left(1-\phi^2\right)\overline{u}_1\,.
\label{eq:lin4} \end{align}
We observe the important feature that the rescaling in Eq.~(\ref{eq:lin3}) 
compensates for the un-conventional normalization for longitudinal component
in Eq.~(\ref{eq:Vdecomp}). That is, (without interactions) $\overline{u}_1$ is the 
wave-function of a single particle state that contributes $\frac{1}{2}\omega$ to 
the VPE. We find that the normalization issue and the construction of an Hermitian 
scattering problem are simply the two sides of the very same medal.

With this scaling the continuity equation is as simple as
$\mu u_0=-\overline{u}_1^\prime$. Using the soliton equations~(\ref{eq:eom}) it is
straightforward to verify that the above fluctuation equations with $\omega=0$ are 
solved by $\eta=\phi^\prime$ and $\overline{u}_1=-\mu a$. The latter relation 
corresponds to $u_0=a^\prime$. Hence this zero mode is nothing but the (infinitesimal) 
translation of the soliton. Observing a zero mode in the bound state spectrum will 
further test the numerical simulations in Chap.~\ref{sec:num}.

For $\mu \ge2$ we can now straightforwardly apply the formalism of Chap.~\ref{sec:real}
with $m_1=2$, $m_2=\mu$ and the potential matrix
\begin{equation}
V=\begin{pmatrix}
6\left(\phi^2-1\right)+2g\phi a^\prime+g^2\left(\phi^2-1\right)^2 
& g\mu\left(1-\phi^2\right) \\[2mm] g\mu\left(1-\phi^2\right) & 0\end{pmatrix}\,.
\label{eq:ProcaPot}\end{equation}
In the other case, $\mu\le2$ we set
$$
D(k)=\imu\begin{pmatrix}k_1 & 0 \cr 0 & k\end{pmatrix}
\qquad {\rm with}\qquad
k_1=k_1(k)=k\sqrt{1-\frac{4-\mu^2}{[k+\imu0^{+}]^2}}
$$
in Eq.~(\ref{eq:JostDEQ}) and replace $m_1\,\to\,\mu$ in Eq.~(\ref{eq:tildeVPE}).

\section{Numerical results}
\label{sec:num}

In this section we present and discuss our numerical results for the VPE of the 
Proca soliton constructed above. We first mention that for all scenarios considered,
we observe an energy eigenvalue in the symmetric channel at around 
$\omega_0\approx0.01\ldots0.03$. This is the translational zero mode. It is not exactly 
at zero because of the discrepancy between the actual soliton profiles and the 
parameterizations in Eq.~(\ref{eq:fit}). This discrepancy provides an additional measure 
for the accuracy of the fit. Minor changes (for example using the fits from $g=1.2$ 
for $g=1.0$) in the fitted profiles fail to produce a low energy bound state at all.

We continue with the comparison of the real and imaginary momentum formalism 
as for the toy model in Sect.~\ref{sec:real}. Six cases are listed in 
Tab.~\ref{tab:VPE_Proca1}. For $\mu=2.5$ there is no contribution associated 
with the Feynman diagram correction in Eq.~(\ref{eq:DiffBorn}) because then 
the self-interaction potential for the heavier particle is zero. The table 
exhibits perfect agreement of the two approaches. Typically we observe 
differences at the fourth significant digit which, however, is out of the 
realm of the numerical precision. Obviously, it is possible to compute the 
VPE with both formalisms and, as expected from the analysis in the previous
chapter, the normalization of the longitudinal component of the vector meson 
field does not hamper the analytic continuation.
\begin{table}
\centerline{
\begin{tabular}{l|lllll|l}
& $\widetilde{E}_{\rm b.s.}$ & $\widetilde{E}_{\rm con.}$
& $\widetilde{E}_{\rm vac}$ & $E_{\rm vac}-\widetilde{E}_{\rm vac}$~~
& $E_{\rm vac}$ & $E_{\rm vac}$\\
\hline
$\mu=1.5\,,g=1.0$ & -0.735 & 0.360 & -0.375 & -0.273 & -0.648 & -0.649\\
$\mu=1.5\,,g=1.2$ & -0.735 & 0.353 & -0.382 & -0.272 & -0.649 & -0.649\\
$\mu=1.5\,,g=1.5$ & -0.757 & 0.371 & -0.387 & -0.266 & -0.653 & -0.653\\
\hline
$\mu=2.5\,,g=1.0$ & -1.146 & 0.459 & -0.687 & 0 & -0.687 & -0.687\\
$\mu=2.5\,,g=1.2$ & -1.164 & 0.453 & -0.711 & 0 & -0.711 & -0.711\\
$\mu=2.5\,,g=1.5$ & -1.172 & 0.440 & -0.732 & 0 & -0.732 & -0.732\\[2mm]
\end{tabular}}
\caption{\label{tab:VPE_Proca1}Comparison of the results from the real 
and imaginary momentum computations of the VPE in the Proca model
for various model parameters. $\widetilde{E}_{\rm b.s.}$, 
$\widetilde{E}_{\rm con.}$ and $\widetilde{E}_{\rm vac}$ respectively 
refer to the bound state, continuum pieces and their sum in Eq.~(\ref{eq:tildeVPE})
while $E_{\rm vac}-\widetilde{E}_{\rm vac}$ is the Feynman diagram
correction from Eq.~(\ref{eq:DiffBorn}). Finally the last column is
the imaginary axis result from Eq.~(\ref{eq:Evac}).}
\end{table}

For the above shown equivalence of the real and imaginary axis calculation,
the fitting functions, Eq.~(\ref{eq:fit}) are good enough. However, for 
more quantitative discussions of the parameter dependence of the VPE a
more ambitious parameterization might be needed. Also, as mentioned after 
Eq.~(\ref{eq:fit}) we directly fit $a^\prime$ because only that part 
of the vector profile enters the differential equation~(\ref{eq:lin4}).
We have considered a number of alternative parameterizations and 
found
\begin{equation}
\phi(x)\approx a_0\tanh(a_1x)+(1-a_0)\tanh(a_2x)
\qquad{\rm and}\qquad
a^\prime(x)\approx \left(b_0+b_1x^2+b_2x^4\right)x{\rm e}^{-b_3x^2}
\label{eq:fit2}\end{equation}
to be most pertinent. We have assessed that from the predicted zero mode energy 
eigenvalue. The closer it is to zero, the more reliable is the considered 
parameterization. In most of the cases, however, there are only minor differences. 
For example, the case $\mu=1.5$ and $g=1.0$ yields $E_{\rm vac}=-0.658$ and $-0.649$ 
for Eqs.~(\ref{eq:fit}) and~(\ref{eq:fit2}), respectively. Generally we must allow 
a parameterization variance of one or two percent.
\begin{table}
\centerline{
\begin{tabular}{l|lllll}
$g$ & 0.4 & 0.8 & 1.2 & 1.6 & 2.0\\
\hline
$\mu=1.0$ &  -0.655 & -0.640 & -0.624 & -0.611 & -0.609\\
$\mu=1.5$ &  -0.662 & -0.648 & -0.657 & -0.663 & -0.682\\
$\mu=2.0$ &  -0.670 & -0.673 & -0.688 & -0.712 & -0.757\\
$\mu=2.5$ &  -0.670 & -0.685 & -0.718 & -0.759 & -0.826\\
$\mu=3.0$ &  -0.673 & -0.696 & -0.737 & -0.804 & -0.890\\[2mm]
\end{tabular}}
\caption{\label{tab:VPE_Proca2}The vacuum polarization energy $E_{\rm VPE}$ for
the Proca soliton as a function of the coupling constant $g$ and for several 
values of the vector meson mass~$\mu$.}
\end{table}

In Tab.~\ref{tab:VPE_Proca2} we present the VPE as a function of the coupling 
constant $g$ as obtained from the imaginary axis formulation, Eq.~(\ref{eq:Evac}). 
After all, we have established its equivalence with the real axis formulation and 
it is much more efficient. Nevertheless we have verified this equivalence for 
selected cases.

When the scalar field is heavier than the Proca field, the VPE shows 
only little dependence on the coupling constant. The VPE is not even
a monotonous function thereof. However, in the other regime, $\mu>2$, the
VPE considerably decreases as the coupling increases. It is a bit surprising
that the Proca model VPE is close to the kink VPE (which in present units 
is $\frac{1}{6}\left(\sqrt{3}-\frac{18}{\pi}\right)\approx-0.666$ \cite{Ra82}) 
when the Proca field is the lighter one because lowering the threshold to $\mu<2$ 
considerably alters the spectrum. In particular, the so-called {\it shape mode} 
bound state in the anti-symmetric channel, which in the kink model is at 
$\omega_1=\sqrt{3}\approx1.732$, may become unbound. Only when the
coupling exceeds a certain value that bound state re-emerges.
This is shown Tab.~\ref{tab:shapemode}. 
\begin{table}
\centerline{
\begin{tabular}{l|lllll}
$g$ & 0.4 & 0.8 & 1.2 & 1.6 & 2.0\\
\hline
$\mu=1.5$ &  -- & -- & 1.499 & 1.429 & 1.330\\
$\mu=2.0$ &  1.720 & $1.684^\ast$ & $1.629^\ast$ & $1.559^\ast$ & $1.478^\ast$\\
$\mu=2.5$ &  $1.724^\ast$ & $1.702^\ast$ & $1.665^\ast$ & $1.615^\ast$ & $1.556^\ast$\\[2mm]
\end{tabular}} 
\caption{\label{tab:shapemode}The bound state energy eigenvalues in the 
anti-symmetric channel as a function of the coupling constant and the 
Proca mass~$\mu$. Entries with a star indicate that there is second
bound state (in addition to the zero mode) in the symmetric channel 
just below threshold at ${\rm min}(2,\mu)$.}
\end{table}
On the other hand, when $\mu\ge2$ we observe a more moderate variation of this 
energy eigenvalue. Yet, the VPE changes considerably as a function of the 
coupling. Hence the change in the bound state spectrum is (partially) compensated
by a similar one of the continuum spectrum. We view this as a manifestation of 
Levinson's theorem which tells us that altering the number of bound
states has a significant impact on the phase shift.

The argument at the end of Chap.~\ref{sec:Proca}, that the kink model would 
be assumed for large $\mu$, however, is not necessarily valid for time 
dependent fluctuations because there is always a frequency $\omega$ such 
that $\mu^2-\omega^2$ is small and we may not generalize the local 
approximation to scattering wave-functions.

\section{Conclusion}
\label{sec:concl}
The main objective of this project has been the investigation of the role
of a potential non-analytic field normalization when computing the 
vacuum polarization energy (VPE) for a soliton containing a massive vector 
meson described by a Proca field. Fortunately it quickly became 
clear that this problem is closely related to the construction of an 
Hermitian scattering problem for the quantum fluctuations about the soliton.
After that construction, the VPE calculation turned into that of two 
coupled scalar fields.

In verifying this conclusion by numerical simulation we have, as an important
byproduct, established the equivalence of the real and imaginary momentum 
formalisms for computing the VPE when there is a mass gap. In this context the 
main accomplishment was to avoid the Born approximation because it is
imaginary for fluctuation energies that are within the mass gap. We have used 
a particular helper function, motivated by the Pauli-Villars regularization 
scheme, and showed analytically as well as numerically that the resulting 
deviation from the no-tadpole condition is compensated by a finite Feynman 
diagram. 

Numerically we have then constructed the soliton in a $D=1+1$ model in which 
a Proca field interacts with a scalar one, solved the wave-equations for the 
small amplitude fluctuations about the soliton, and extracted the Jost function, 
both for real and purely imaginary momenta. This function is central to the 
spectral methods approach to compute the VPE. These methods are particularly 
efficient when the fluctuation momenta are continued to the imaginary axis.

While the classical energy increases with the coupling constant, the VPE 
either only varies mildly when the Proca field is the lighter of the two 
fields or decreases considerably with the coupling strength when the Proca 
field is the heavier one. A qualitative comparison of classical and quantum 
contributions to the energy would only be possible if the overall factor 
of the Lagrangian that acts as a loop-counter and emerges by scaling fields 
and coordinates to dimensionless quantities, was known.

Here we have considered the simplest model producing a massive vector meson soliton. 
Eventually we should consider higher dimensions and/or allow multiple scalar fields. 
The ultimate goal is the computation of the VPE of the 't Hooft-Polyakov 
monopole~\cite{tHooft:1974kcl,Polyakov:1974ek}. Though numerically subtle, the heavy 
Proca mass limit is interesting because with the derivative coupling in 
Eq.~(\ref{eq:lag1}) it potentially induces a non-trivial coefficient function 
for the kinetic term of the scalar field. A scenario for which a number of
solitons have recently been constructed, {\it cf.\@} Ref.~\cite{daHora:2024nmt}
which also quotes many articles that discuss models with such solitons.
It seems infeasible to directly compute the VPE in such models because the
wave-equations are not of the form, Eq.~(\ref{eq:toy1}) as relative factors
between the time and space derivatives may emerge.

\acknowledgments{D.\@ A.\@ P.\@ acknowledges support by a bursary from the 
National Research Foundation of South Africa (NRF) and H.\@ W.\@ is supported in 
part by the NRF under grant~150672.}

\end{document}